# Spin Wave Interference Detection via Inverse Spin Hall Effect

Michael Balynskiy, Howard Chiang, David Gutierrez, and Alexander Khitun[*]

*Department of Electrical and Computer Engineering, University of California -Riverside, Riverside, California, USA 92521*

**Abstract**

In this letter, we present experimental data demonstrating spin wave interference detection using spin Hall effect (ISHE). Two coherent spin waves are excited in a yttrium-iron garnet (YIG) waveguide by continuous microwave signals. The initial phase difference between the spin waves is controlled by the external phase shifter. The ISHE voltage is detected at a distance of 2 mm and 4 mm away from the spin wave generating antennae by an attached Pt layer. Experimental data show ISHE voltage oscillation as a function of the phase difference between the two interfering spin waves. This experiment demonstrates an intriguing possibility of using ISHE in spin wave logic circuit converting spin wave phase into an electric signal.

Corresponding Author: akhitun@engr.ucr.edu

There is a big impetus in the development of spin-based logic devices aimed to benefit spin in addition to charge [1]. Spin wave (SW) devices are one of the promising approaches exploiting collective spin oscillations [2]. SW can propagate on much larger distances (e.g., 1 cm at room temperature) compared to the spin-diffusion length in metals [3]. It makes it possible to exploit the phase of SW signal for information transfer [4]. There were several prototypes demonstrated during the past decade including SW majority logic gate [5], SW holographic memory [6], and devices for special type data processing [7, 8]. In all of the above cited works, SW output was detected by micro-antenna which implies certain restriction on the size/position of the input-output ports due to the stray field coupling. ISHE is one of the possible alternatives which may be more scalable and less sensitive to the direct input-output coupling. SW detection using ISHE was presented in a number of works [9-12]. For instance, it was experimentally demonstrated magnon spin transport between the spatially separated inductive pulse spin-wave source and the ISHE detector [12]. The role of the travelling SW was revealed by a delay in the detection of the ISHE voltage. In this letter, we extend this approach to two-SW interference. The primarily objective of this work is to demonstrate the correlation between the phase difference of the interfering spin waves and the produced ISHE voltage.

The test structure is schematically shown in Fig. 1. It comprises a 9.5 µm thick YIG waveguide (10 mm × 2 mm) with a 10 nm thick (0.2 × 2 mm$^2$, 374 Ohm) Pt strip deposited on the top. The YIG waveguide is magnetized along its long axis by an external bias magnetic field providing conditions for the propagation of a backward volume magnetostatic wave (BVMSW). There are two 50 µm-wide Au microstrip antennae fabricated on the edges of waveguide. These antennae are used for continuous SW signal generation. The antennas are intentionally placed at the different distances from the Pt detector. The distance from the left antenna to the center of the detector is about 2 mm while the distance from the antenna on the right to the detector is 4 mm. The asymmetry in the excitation port position helps to unmask the effect of direct coupling on ISHE voltage. The antennae are connected to a programmable network analyzer (PNA) Keysight N5241A. PNA serves as a common input for the both antennae. It also allows us to detect SW signal independently from ISHE measurements. There is a phase shifter included in the antennae-PNA circuit. This shifter is to control the initial phase difference between the excited spin waves. The DC ISHE voltage is detected via Lock-In Amplifier SR830 DSP which is synchronized with sweeping frequency of PNA operating in power sweep mode.

The test structure is placed inside an electromagnet (GMW model 3472 – 70) with the pole cap of 50 mm (2 inch) diameter tapered. The system provided a uniform bias magnetic field $\Delta H/H < 10^{-4}$ per 1 mm in the range from −2000 Oe to +2000 Oe. Based on the power source specification (KEPCO), the magnetic field instability was estimated about 0.15 Oe. All measurements are accomplished at room temperature.

The first set of experiments is aimed at confirming the spin wave generation by the antennas via the inductive voltage measurements, and finding the optimum operational frequency $f$ and at the bias magnetic field $H_0$. Based on our prior studies of SW propagation in YIG films [13, 14], we searched for the maximum output signal in the frequency range from 4 GHz to 6 GHz. In Fig.2., there are shown experimental data for $S_{21}$ parameter in the loop PNA – antenna 1 – antenna2 – PNA. The maximum signal is observed for *f = 4.972 GHz* and bias magnetic field $H_0 = 1100$ Oe. This combination of frequency and bias magnetic field fits well the BVMSW dispersion given by [15]:

$$f_{BVMSW} = \sqrt{f_H \left(f_H + f_M \frac{1-\exp(-kd)}{kd}\right)}, \tag{1}$$

where $f_H = \gamma H_0$, $f_M = 4\pi\gamma M_0$, $\gamma = 2.8\ MHz/Oe$, $k \sim 30\ cm^{-1}$ is the SW wavenumber, $d$ is the film thickness. The data show prominent spin wave propagation over the 10 mm distance at room temperature.

Next, ISHE voltage produced by spin waves was detected across the Pt strip. Spin waves are excited by the two antennae biased by PNA with the fixed frequency *f = 4.972 GHz*. The attenuators (e.g. A1 and A2 shown in Fig.1) were used to equalize the amplitudes of SW signals coming to Pt detector. ISHE voltage was measured at different bias magnetic fields. The experimental data are shown in Fig. 3. In Fig.3(a) there is shown the dependence for magnetic field directed along the axes of YIG waveguide as shown in Fig.1. The maximum of the ISHE voltage $-5.5\ \mu V$ appears at $H_0 = 1100$ Oe which corresponds to the maximum of SW signal. In order to further verify the origin of the detected voltage, the direction of the bias magnetic field has been reversed. The electric field induced by the ISHE $E_{ISHE}$ can be written as follows [16]:

$$\vec{E}_{ISHE} \propto \vec{J}_s \times \vec{\sigma}, \tag{2}$$

where $J_s$ is the spin current injected from YIG in Pt and $\sigma$ is the spin polarization vector of the spin current defined by the bias magnetic field $H_0$. As expected from Eq.(2), the reverse of the bias magnetic field resulted in the ISHE voltage polarity change as can be seen in Fig. 3(b). The maximum of the detected voltage 4.2 µV appears at $H_0 = -1100$ Oe. A difference in the peak ISHE voltage seen in Fig.3(a) and Fig.3(b) can be attributed to the weak non-reciprocity of BVMSW and to an asymmetrical orientation of our device in magnetic field. The collection of experimental data in Figs.2 and 3 confirms the origin of the ISHE voltage as a result of spin wave conversion into an electron carried spin current.

Finally, ISHE voltage was measured as a function of the phase difference between the propagating spin waves. Spin waves were generated by antenna 1 and antenna 2 at constant *f = 4.972 GHz*. The bias magnetic field was fixed to $H_0 = 1100$ Oe. The initial phase difference between the antennae is the only parameter which has been changed. The obtained experimental data are shown in Fig.4. It is clearly seen the oscillation of the ISHE voltage depending on the phase difference of interfering spin waves. The maximum voltage is $-5.5\ \mu V$ which is the same as in Fig.3(a). The minimum voltage is about $-0.6\ \mu V$. This minimum voltage is attributed to the destructive SW interference corresponding to the phase difference $\Delta\phi = \pi$. All other phases in Fig.4 are defined to this phase difference. The oscillation of the ISHE voltage can be well approximated by the classical formula:

$$V_{ISHE} = \sqrt{V_1^2 + V_2^2 + 2V_1V_2\cos(\Delta\phi)} \qquad (3)$$

where $V_1$ and $V_2$ are the voltages produced separately by each SW. There may be several reasons for $V_{ISHE}$ not coming to zero for the destructive spin wave interference (i.e., $\Delta\phi = \pi$). For instance, the wave front of the SW signals may be disturbed by the reflection from the waveguide boundaries.

There are several observations we want to make based on the obtained experimental data. (i) ISHE voltage is produced by the interfering spin waves. This conclusion is supported by the experimental data shown in Fig.2 and Fig.3. ISHE voltage attains its maximum for magnetic field allowing BVMSW propagation which is in good agreement with the results of PNA measurements (e.g., more than 10 dB increase of $S_{21}$ parameter). The combination of frequency and bias magnetic field is well fitted by the BVMSW dispersion (i.e., Eq. 1). (ii) There is an ISHE voltage dependence on the phase difference between the interfering spin waves. The voltage oscillates with the phase difference following the classical interference formula Eq. 3. There is a prominent difference between the maximum and minimum voltages which is attributed to the cases of constructive and destructive SW interference, respectively. A careful comparison of the ISHE voltage measured across Pt strip and SW intensity measured via PNA shows that the maximum of the DC signal occurs for constructive spin wave interference.

In conclusion, we presented experimental data demonstrating the detection of two interfering spin waves via ISHE voltage measurements. The amplitude of the voltage oscillates depending on the phase difference between the waves. ISHE detectors may be potentially utilized for connecting phase-based SW logic devices with output electronic devices [6]. It may be also possible to remotely control pure spin currents using spin waves interference.

**Acknowledgement**

This was supported in part by the National Science Foundation (NSF) under Award # 2006290. It was also supported in part by the Spins and Heat in Nanoscale Electronic Systems (SHINES), an Energy Frontier Research Center funded by the U.S. Department of Energy, Office of Science, Basic Energy Sciences (BES) under Award # SC0012670.

*The data that support the findings of this study are available from the corresponding author upon reasonable request.*

**Figure Captions**

**Figure 1:** Schematic illustration of the experimental setup: two coherent spin waves are excited in the YIG waveguide using two microstrip antennae. The antennae are connected to PNA via the set of attenuators (A1 and A2) and a phase shifter. There is Pt stripe on top of the waveguide for ISHE voltage detection. The distances from the strip to the antennae are 2 mm and 4 mm.

**Figure 2:** Experimental data showing the coupling (i.e., S21 parameter) between the antennae depending on the bias magnetic field $H_0$. The coupling is maximum at $H_0 = 1100$ Oe for excitation frequency *f = 4.972 GHz*, which is attributed to BVMSWs propagating in the waveguide.

**Figure 3:** Experimental data showing the variation of the ISHE voltage depending on the bias magnetic field. (a) Magnetic field is directed along the axes of YIG waveguide. b)The direction of magnetic field is reversed.

**Figure 4:** Experimental data showing ISHE voltage oscillation depending on the phase difference between the interfering spin waves.

**References**

[1] S. A. Wolf, A. Y. Chtchelkanova, and D. M. Treger, "Spintronics - A retrospective and perspective," *Ibm Journal of Research and Development,* vol. 50, pp. 101-110, Jan 2006

[2] A. Khitun and K. L. Wang, "Nano scale computational architectures with Spin Wave Bus," *Superlattices and Microstructures,* vol. 38, pp. 184-200, Sep 2005. DOI: 10.1016/j.spmi.2005.07.001

[3] J. Bass and W. P. Pratt, "Spin-diffusion lengths in metals and alloys, and spin-flipping at metal/metal interfaces: an experimentalist's critical review," *Journal of Physics: Condensed Matter,* vol. 19, 2007

[4] A. Khitun, "Spin wave phase logic," *Cmos and Beyond: Logic Switches for Terascale Integrated Circuits,* pp. 359-378, 2015

[5] Y. Wu, M. Bao, A. Khitun, J.-Y. Kim, A. Hong, and K.L. Wang, " A Three-Terminal Spin-Wave Device for Logic Applications," *Journal of Nanoelectronics and Optoelectronics,* vol. 4, pp. 394-7, 2009

[6] F. Gertz, A. Kozhevnikov, Y. Filimonov, and A. Khitun, "Magnonic Holographic Memory," *Ieee Transactions on Magnetics,* vol. 51, Apr 2015. DOI: 10.1109/tmag.2014.2362723

[7] A. Kozhevnikov, F. Gertz, G. Dudko, Y. Filimonov, and A. Khitun, "Pattern recognition with magnonic holographic memory device," *Applied Physics Letters,* vol. 106, Apr 2015. DOI: 10.1063/1.4917507

[8] Y. Khivintsev, M. Ranjbar, D. Gutierrez, H. Chiang, A. Kozhevnikov, Y. Filimonov, and A. Khitun, "Prime factorization using magnonic holographic devices," *Journal of Applied Physics,* vol. 120, Sep 2016. DOI: 10.1063/1.4962740

[9] K. Ando, J. Ieda, K. Sasage, S. Takahashi, S. Maekawa, and E. Saitoh, "Electric detection of spin wave resonance using inverse spin-Hall effect," *Appl. Phys. Lett.,* vol. 94, 2009. DOI: https://doi.org/10.1063/1.3167826

[10] L. Feiler, K. Sentker, M. Brinker, N. Kuhlmann, F.-U. Stein, and G. Meier, "Inverse spin-Hall effect voltage generation by nonlinear spin-wave excitation," *Phys. Rev. B,* vol. 93, 2016. DOI: https://doi.org/10.1103/PhysRevB.93.064408

[11] T. Brächer, M. Fabre, T. Meyer, T. Fischer, S. Auffret, O. Boulle, U. Ebels, P. Pirro, and G. Gaudin, "Detection of Short-Waved Spin Waves in Individual Microscopic Spin-Wave Waveguides Using the Inverse Spin Hall Effect," *Nano Letters,* vol. 17, pp. 7234–7241, 2017. DOI: https://doi.org/10.1021/acs.nanolett.7b02458

[12] A. V. Chumak, A. A. Serga, M. B. Jungfleisch, R. Neb, D. A. Bozhko, V. S. Tiberkevich, and B. Hillebrands, "Direct detection of magnon spin transport by the inverse spin Hall effect," *Applied Physics Letters,* vol. 100, 2012. DOI: https://doi.org/10.1063/1.3689787

[13] A. Kozhevnikov, F. Gertz, G. Dudko, Y. Filimonov, and A. Khitun, "Pattern recognition with magnonic holographic memory device," *Applied Physics Letters,* vol. 106, p. 142409, Apr 6 2015. DOI: 14240910.1063/1.4917507

[14] M. Balynsky, A. Kozhevnikov, Y. Khivintsev, T. Bhowmick, D. Gutierrez, H. Chiang, G. Dudko, Y. Filimonov, G. X. Liu, C. L. Jiang, A. A. Balandin, R. Lake, and A. Khitun, "Magnonic interferometric switch for multi-valued logic circuits," *Journal of Applied Physics,* vol. 121, p. 024504, Jan 2017. DOI: 10.1063/1.4973115

[15] A. A. Serga, A. V. Chumak, and B. Hillebrands, "YIG magnonics," *Journal of Physics D-Applied Physics,* vol. 43, Jul 7 2010. DOI: 264002 10.1088/0022-3727/43/26/264002

[16] V. E. Demidov, M. P. Kostylev, K. Rott, P. Krzysteczko, G. Reiss, and S. O. Demokritov, "Excitation of microwaveguide modes by a stripe antenna," *Appl. Phys. Lett.,* vol. 95, 2009. DOI: https://doi.org/10.1063/1.3231875

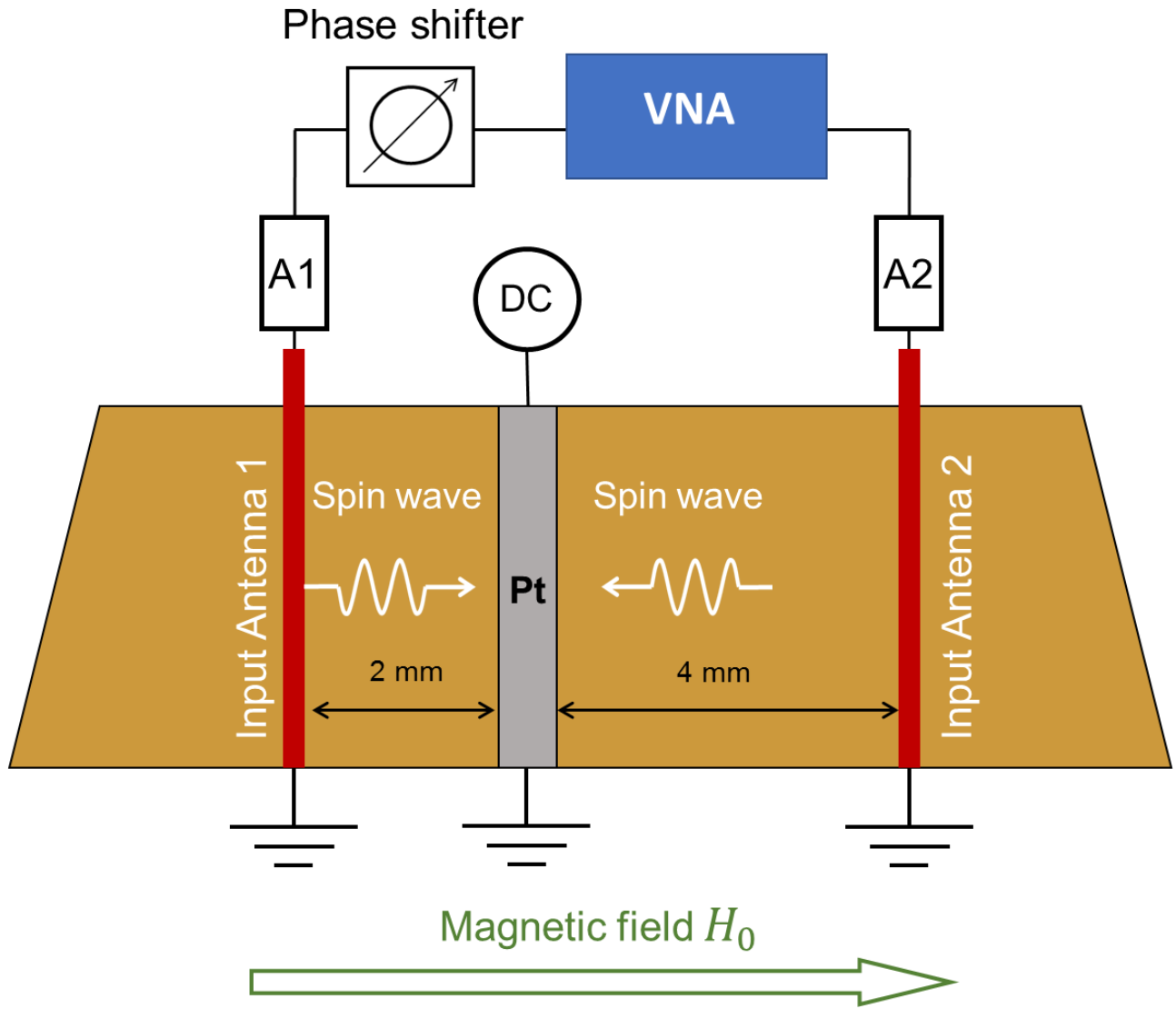

**Figure 1**

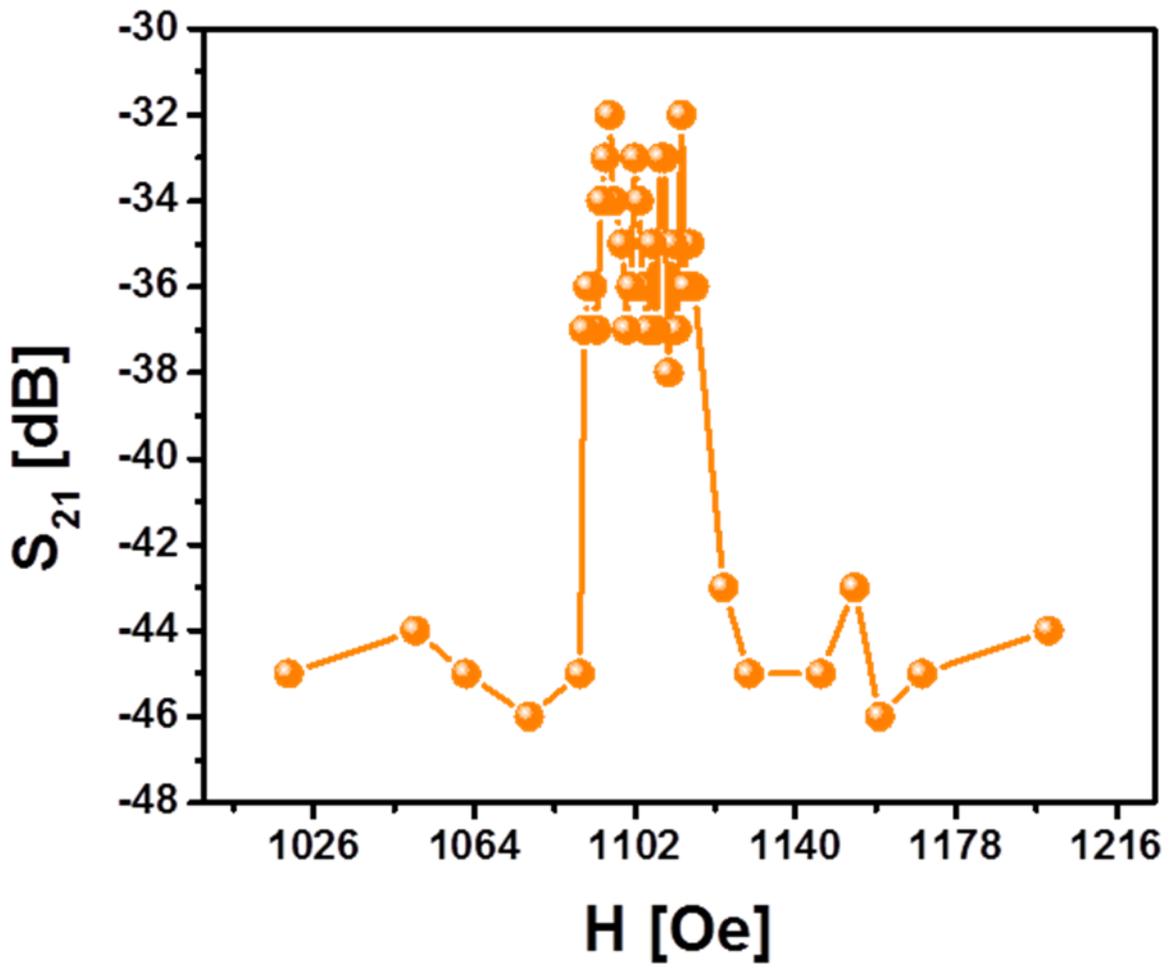

**Figure 2**

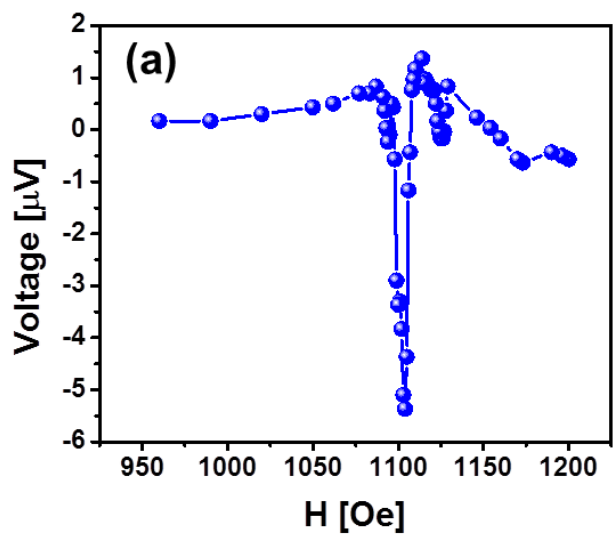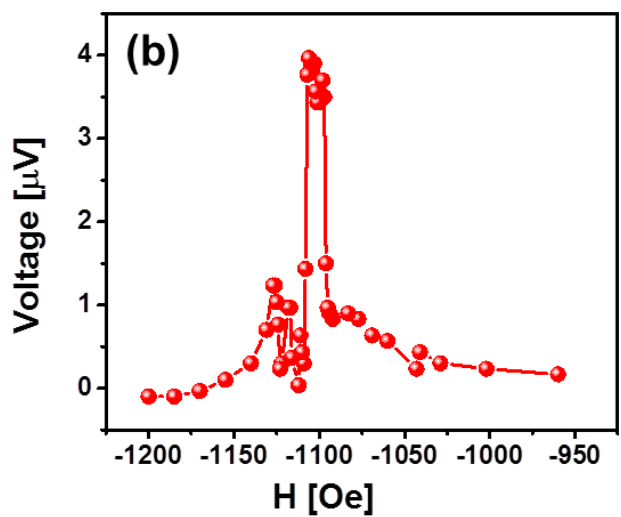

**Figure 3**

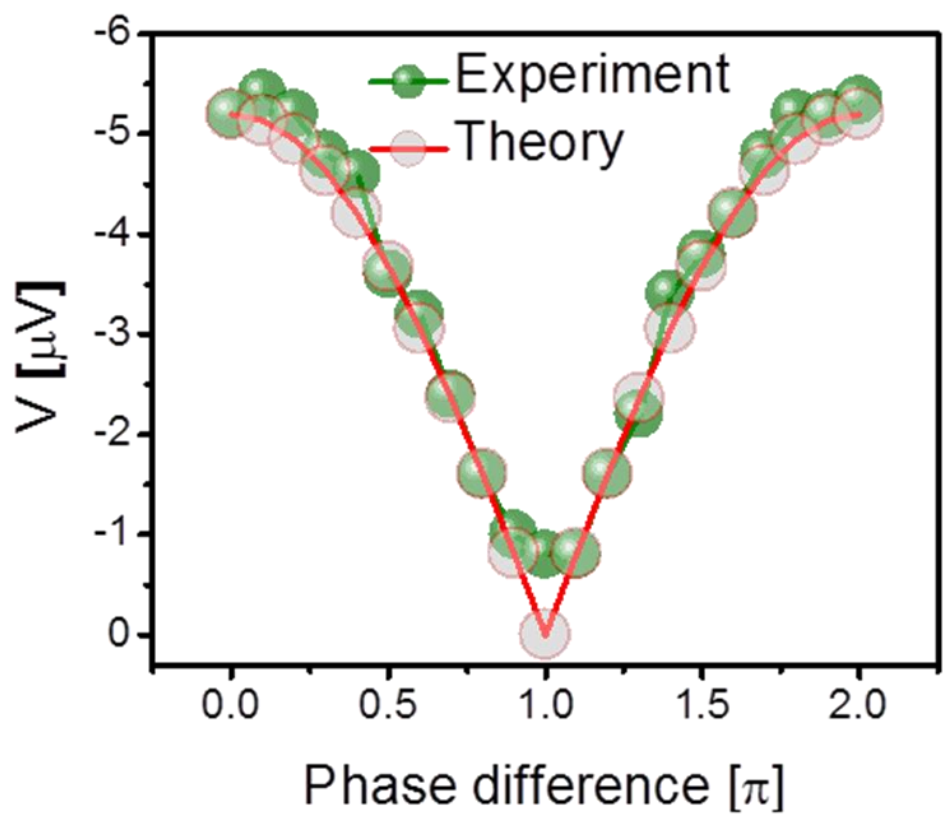

**Figure 4**